\documentclass[%
 aip,
 amsmath,amssymb,
 reprint,%
]{revtex4-1}
\usepackage{graphicx}
\usepackage{dcolumn}
\usepackage{enumitem}
\usepackage{mathtools}
\usepackage{xcolor}
\usepackage[normalem]{ulem}
\usepackage{comment}
\usepackage{bbm}
\usepackage{bm}

\usepackage[utf8]{inputenc}
\usepackage[T1]{fontenc}
\usepackage{etoolbox}
\usepackage{array}
\usepackage{booktabs}
\usepackage{multirow}

\usepackage[breaklinks=true]{hyperref}
\usepackage{breakcites}

\makeatletter
\def\@email#1#2{%
 \endgroup
 \patchcmd{\titleblock@produce}
  {\frontmatter@RRAPformat}
  {\frontmatter@RRAPformat{\produce@RRAP{*#1\href{mailto:#2}{#2}}}\frontmatter@RRAPformat}
  {}{}
}%

\makeatother

\preprint{AIP/123-QED}

\begin{document}

\title{Hybrid Monte Carlo Metadynamics (hybridMC-MetaD)}

\author{Charlotte Shiqi Zhao}
\affiliation{Department of Chemical Engineering, University of Michigan, Ann Arbor, MI 48109, USA\looseness=-1}

\author{Sun-Ting Tsai}
\affiliation{Department of Chemical Engineering, University of Michigan, Ann Arbor, MI 48109, USA\looseness=-1}

\author{Sharon C. Glotzer}
    \email{sglotzer@umich.edu}
\affiliation{Department of Chemical Engineering, University of Michigan, Ann Arbor, MI 48109, USA\looseness=-1}
\affiliation{Biointerfaces Institute, University of Michigan, Ann Arbor, MI 48109, USA}

    \date{\today}

\begin{abstract}
We propose the powerful integration of the Hybrid Monte Carlo (hybridMC) algorithm and Well-Tempered Metadynamics. This new algorithm, hybridMC-MetaD, enhances the flexibility and applicability of metadynamics by allowing for the utilization of a wider range of collective variables (CVs), namely non-differentiable CVs.
We demonstrate the usage of hybridMC-MetaD through five examples of rare events in molecular dynamics (MD) simulations, including a rare transition in a model potential system, condensation of the argon system, crystallization in a nearly-hard sphere system, a nearly-hard bipyramid system and a colloidal suspension. By taking advantage of hybridMC, which combines molecular dynamics (MD) and MC, we are able to bias the transitions along non-differentiable CVs for all five cases, which would be unfeasible with conventional MD simulations. Enabled by metadynamics, we observed significant acceleration of the phase transitions and calculated free energy barriers using the hybridMC-MetaD simulation data. For the nearly-hard bipyramid system whose crystallization is primarily driven by entropy, we report the free energy surface for the first time. Through our case studies, we show that our hybridMC-MetaD scheme reduces the complexity of using metadynamics and increases its accessibility. We believe the hybridMC-MetaD algorithm will stimulate greater interest in, and foster broader applications of metadynamics.
\end{abstract}

\maketitle

\section{Introduction}
\label{sec:introduction}

Rare events are ubiquitous in nature across length scales, contrary to what the name might suggest. Examples include chemical reactions, nucleation of droplets and crystals, protein folding and aggregation, and earthquakes. The rarity of these events stems from infrequent and stochastic fluctuations large enough to trigger subsequent changes in the system states \cite{hussain2020studying, allen2006simulating}. Thermodynamically, these critical fluctuations are infrequent because their occurrence requires overcoming free energy barriers. As a result, a feature of rare events is the separation of time scales. More specifically, the average wait time between consecutive rare events is much longer than the duration of the event, and during the wait time, the system dwells in a basin of metastable states separated by free energy barriers. Due to the time scale issue and stochasticity, observing rare events in computer simulations requires prolonged run times and sensitive detection techniques \cite{butler2024change}.
Therefore, conventional simulations tend to waste resources sampling metastable states instead of the relevant rare event, which may not even occur within a feasible simulation time \cite{hussain2020studying, allen2006simulating, yang2019enhanced, henin2022enhanced}. To improve sampling efficiency, numerous enhanced sampling techniques have been developed \cite{henin2022enhanced, yang2019enhanced}. These methods have enabled mechanistic studies of a variety of importance processes \cite{shen2023enhanced}, such as the kinetics of protein-drug unbinding \cite{tiwary2015kinetics} and the competing melting pathways of copper and aluminum \cite{samanta2014microscopic}.

One particular class of enhanced sampling methods, which includes, e.g., umbrella sampling \cite{torrie1977nonphysical} and metadynamics \cite{laio2002escaping}, utilizes a bias potential to temper the high free energy barrier. In these methods, the bias potential is applied along one or more collective variable(s) (CVs), which are lower dimensional representations of the system coordinates. Examples of CVs include dihedral angles, cluster sizes, and structural order parameters. A careful choice of CV is essential to the success of these methods. However, identifying a suitable CV can be challenging as we may have only very limited understanding of the systems under investigation \cite{rohrdanz2013discovering}. For instance, CVs should adopt distinct values for different metastable states. But for a polymorphic system, it can be especially difficult to find one rudimentary CV that has different values for all the relevant polymorphs. As such, many research efforts have focused on developing CVs for various systems and processes \cite{tiwary2016spectral, wang2021state, bonati2023unified}.

Ideally, the choice of CV should be guided solely by its effectiveness in capturing the dynamics of the system and its ability to distinguish the relevant states along the dynamical transition pathways. In practice, however, the calculation of bias forces requires CVs to be continuous and differentiable with respect to the system coordinates \cite{fiorin2013using, meraz2024simulating}. Therefore, additional switching functions are needed for integer-valued CVs, such as the coordination number \cite{bonomi2009plumed}. This requirement especially hinders the adoption of CVs with complex mathematical formulae, as it falls on the user to calculate and implement the derivatives of the CVs \cite{zhang2023artificial}. Automatic differentiation frameworks \cite{paszke2019pytorch, abadi2016tensorflow, frostig2018compiling} provide a remedy, but integrating those frameworks with enhanced sampling methods can still be a daunting task.

In Hybrid Monte Carlo (hybridMC) \cite{duane1987hybrid}, the MC importance sampling algorithm is used to accept or reject trial configurations generated by MD. Inspired by several works in which hybridMC is used in combination with umbrella sampling (Refs. \citenum{limmer2011putative, geiger2013neural, gonzalez2014nucleation, peters2017273, gispen2024variational}, to list a few), we propose the hybridMC-MetaD algorithm, in which we couple hybridMC with well-tempered metadynamics \cite{barducci2008well}. Using a modified acceptance criterion, we eliminate the need to explicitly incorporate the bias potential in MD. By doing so, we avoid the evaluation of the bias forces and hence the calculation of the derivatives of the CVs but still take full advantage of metadynamics. As we will demonstrate, the hybridMC-MetaD algorithm enables the usage of a wider range of CVs and we believe it is a more accessible and easier-to-implement option for using metadynamics that will motivate broader applications of metadynamics.

Our paper begins with the presentation of the hybridMC-MetaD algorithm in Sec.~\ref{sec:theory}, along with heuristics on selecting the associated parameters. Next, we discuss in Sec.~\ref{sec:method} the details of how we performed hybridMC-MetaD simulations for five systems and introduce the CVs we used for each system. In Sec.~\ref{sec:results}, we report our results. We first validate our approach on two systems, a model potential system with only two degrees of freedom and the argon system \cite{tsai2019reaction}, by reproducing the known energy barriers. Then we apply our hybridMC-MetaD algorithm to the following crystallization processes: crystallization of a nearly-hard sphere system \cite{filion2011simulation} at two densities, at one of which the process is two-step, a two-step crystallization pathway in a DLVO colloidal suspension \cite{finney2023variational}, and crystallization of a nearly-hard bipyramid system. The transitions in these systems are all significantly accelerated by hybridMC-MetaD, and the improved sampling allows us to calculate the free energy barrier in each case. For the DLVO system, we construct a two-dimensional free energy surface with five metastable states. For the nearly-hard bipyramid system, we observe back-and-forth transitions between the fluid phase and the crystal phase in our hybridMC-MetaD simulations, and we present, for the first time, the free energy surface of this system whose crystallization is driven by entropy.

\section{Theoretical background}
\label{sec:theory}
In this section, we first briefly review the basics of well-tempered metadynamics and the hybridMC algorithm. Then we describe our proposed hybridMC-MetaD algorithm in detail.

\subsection{Standard well-tempered metadynamics}
\label{sec:metad}
Metadynamics \cite{laio2002escaping} utilizes a history-dependent Gaussian bias potential $V_{\rm bias}(\textbf{s})$ to enhance the fluctuations of the biased CV and encourage the exploration of less favored metastable configurations. Metadynamics has been shown to be an efficient method for investigating free energy landscapes and calculating kinetic rates \cite{ray2023kinetics}.  Here, we use well-tempered metadynamics in which the height of the Gaussian bias potential is adaptively tempered to facilitate convergence \cite{barducci2008well}. The bias potential is usually constructed in an iterative manner and can be updated as follows:
\begin{align}
    \begin{aligned}
        V(\textbf{s}, t+1) &= V(\textbf{s}, t) + W_t\exp\left[-\frac{\|\textbf{s}-\textbf{s}_t\|^2}{2\sigma_{G}^2}\right], \\
        W_t &= \omega e^{-V(\textbf{s}_t) / \Delta T},
        \label{eq:bias_update}
    \end{aligned}
\end{align}
where $\textbf{s}_t = \textbf{s}(\mathbf{R}_t)$ is the CV, and is usually defined as a differentiable function of a system's configuration $\mathbf{R}$ in order to compute the resulting forces from the bias potential. The need for continuous and differentiable CVs often necessitates more sophisticated CV design. By leveraging hybridMC, we can alleviate this differentiability restriction and accommodate a wider range of CVs.

\subsection{Standard hybridMC algorithm}
\label{sec:hybridMC}

In this section, we review the hybridMC method \cite{duane1987hybrid, mehlig1992hybrid, guo2018hybrid, inagaki2022hybrid}.

\begin{figure}[!t]
    \includegraphics[width=0.85\columnwidth]{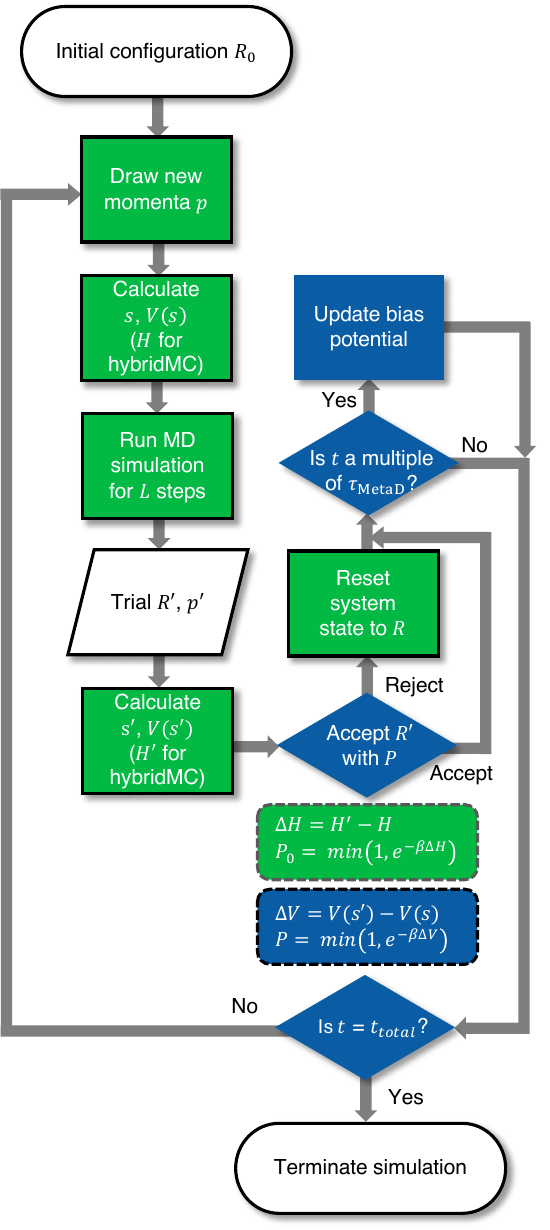}
    \caption
    {\textbf{Flowchart of the hybridMC-MetaD algorithm}. Starting from an initial configuration of the system $R_{0}$, we draw new particle momenta from the Maxwell-Boltzmann distribution $P(p_{i}) \propto \exp{(-\beta p_{i}^2/2m_{i})}$ and calculate the total energy $H$, collective variable $s$, and an initial bias potential $V(s)$. Then we evolve the system by performing a short MD simulation for $L$ steps using a symplectic integration scheme such as the velocity Verlet algorithm to generate a trial configuration $R'$. Next, we calculate the total energy ($H'$), collective variable ($s'$) and bias potential ($V(s')$) associated with $R'$. For an unbiased hybridMC simulation, we accept $R'$ with the acceptance probability ${\displaystyle P_0 = \min{(1, e^{-\beta \Delta H})}}$; for our hybridMC-MetaD algorithm, we accept $R'$ with ${\displaystyle P = \min{(1, e^{-\beta \Delta V})}}$.
    }\label{fig:algorithm}
\end{figure}

hybridMC refers to a type of Markov chain Monte Carlo method in which trial moves are generated by a short molecular dynamics (MD) trajectory. As described in Fig.~\ref{fig:algorithm}, a typical hybridMC algorithm consists of the following steps:
(a) Draw particle momenta from the Maxwell-Boltzmann distribution: 
\begin{align}
    P(\mathbf{v}) = \prod_{i = 1}^{N}(\frac{\beta m_{i}}{2\pi})^{(3/2)} e^{-\frac{\beta}{2}m_{i}(\mathbf{v}_{i}  \cdot \mathbf{v}_{i} )},
    \label{eq:mb}
\end{align}
where $\beta = (k_{B} T)^{-1}$, $N$ is the number of particles in the system, $\mathbf{v}_{i}$ and $m_{i}$ are the velocity vector and mass, respectively, of particle $i$.
(b) Evolve the system using a symplectic integrator such as leapfrog or velocity Verlet for a short period of time ($L$ timesteps).
(c) Accept the resulting configuration with the probability
\begin{align}
    \begin{aligned}
        P 
        & = \min\left( 1, e^{-\beta \Delta H}\right),
        \label{eq:hybridMC_metropolis}
    \end{aligned}
\end{align}
where $\Delta H = H' - H$ is the change in the Hamiltonian of the system following the short MD simulation in step (b).

\subsection{hybridMC-MetaD algorithm}
\label{sec:algorithm}

Our proposed hybridMC-MetaD method modifies the hybridMC method only slightly.

It can be easily shown that if a bias potential $V(\mathbf{s},t)$ is applied to a hybridMC simulation, we need only to modify Eq.~\ref{eq:hybridMC_metropolis} to get a new acceptance rule:
\begin{align}
    P = \min\left( 1, e^{-\beta [\Delta H + \Delta V]} \right).
    \label{eq:hybridMC_bias_metropolis}
\end{align}
Note that both $H$ and $V=V(\mathbf{s},t)$ are dependent on $t$. In particular, it can be further shown \cite{gonzalez2014nucleation, peters2017273} that the $\Delta H$ in Eq.~\ref{eq:hybridMC_bias_metropolis} can be omitted if we use a thermostat such as the Bussi-Donadio-Parrinello (BDP) thermostat \cite{bussi2007canonical} to keep the system's temperature constant.
We present the derivation in the \hyperref[app:derivation]{Appendix}, which results in an acceptance probability that depends only on the change in the bias potential:
\begin{align}
    P = \min\left( 1, e^{-\beta\Delta V} \right).
    \label{eq:bmc_bias_metropolis}
\end{align}
Importantly, this means that in a hybridMC-MetaD simulation, we do not need to incorporate the bias potential explicitly in the short MD simulations that generate the trial configurations. Instead, we simply modify the acceptance rule to sample the biased ensemble (Fig.~\ref{fig:algorithm}). This simplification allows us to bypass the calculation of the derivatives of the CVs and thus removes the inconvenient constraint that CVs be continuous and differentiable.

To reweight the hybridMC-MetaD simulation data and reconstruct the free energy profile, we follow the standard metadynamics calculation detailed in Ref. \citenum{tiwary2015time}.

\section{Methods}
\label{sec:method}
In this section, we first discuss how we choose parameters for hybridMC-MetaD simulations. Then we describe the simulation details for the systems considered in this work. Finally, we introduce our choice of CVs for each system.

\subsection{hybridMC-MetaD parameters}
\label{sec:params}

We summarize the parameters one needs to choose for the hybridMC-MetaD algorithm here and provide heuristics on how to make reasonable choices.

In the standard hybridMC algorithm, the size of the timestep, $dt$, and the length of the trial MD simulation, $L$, can be tuned to optimize sampling efficiency \cite{peters2017273}. Larger $dt$ and $L$ lead to larger trial moves, which can accelerate sampling but may also result in larger errors in energy and thus a lower acceptance probability according to Eq.~\ref{eq:hybridMC_metropolis}. For hybridMC-MetaD, we do not need to use large $dt$ or $L$, as we leverage metadynamics to enhance sampling efficiency and accelerate the transitions. As shown in Ref. \citenum{gonzalez2014nucleation}, with lower values of $dt$, the acceptance probability for hybridMC simulations is close to $100\%$, and the evaluation of the Metropolis criterion may even be omitted. Indeed, in our preliminary tests without the bias potential, the acceptance probability, which is the number of accepted configurations divided by the total number of trial configurations, is also close to $100\%$ for the $dt$ values we selected (listed in Table~\ref{tab:params}). In the case of hybridMC-MetaD, however, the acceptance probability is greatly affected by the metadynamics parameters, especially the choice of the initial height of the bias potential, $\omega_{0}$. We find that with an appropriate choice of $\omega_{0}$, the acceptance probability is usually around or above $70\%$ when the system is exploring the initial metastable state, and increasing $\omega_{0}$ leads to a decrease in the acceptance probability. Moreover, the acceptance probability varies throughout the simulation due to changes in the bias potential height, and it is often at its lowest ($20\% - 40 \%$) when the transition occurs, as the bias potential height typically peaks at this stage. 

In conventional metadynamics, the bias potential is calculated at every timestep $dt$. However, in hybridMC-MetaD, the bias potential is only calculated each time we evaluate the modified Metropolis criterion (Eq.~\ref{eq:bmc_bias_metropolis}), i.e., every $Ldt$ timesteps. Therefore, we define the bias potential update interval, $\tau_{\text{MetaD}}$, for our algorithm (also referred to as ``deposition interval'' in the literature \cite{barducci2008well}) as a multiple of $Ldt$. In other words, the bias potential height is updated using Eq.~\ref{eq:bias_update} every $k \cdot Ldt$ timesteps, where $k$ is an integer. Empirical rules on how to choose the bias potential update interval and other metadynamics parameters (initial height $\omega_{0}$ and width $\sigma_{G}$ of the Gaussian kernel and the bias factor $\gamma$) are discussed in several review papers, such as Refs. \citenum{valsson2016enhancing, bussi2015free}.

\subsection{Simulation details}
\label{sec:sim}

All the simulations for this work were performed with the open-source simulation package HOOMD-blue \cite{anderson2020hoomd} (version $4.7.0$). We used our hybridMC-MetaD algorithm to run all the biased simulations and conventional MD scheme for the unbiased ones. We explain the details of both types of simulations in this section.

\begin{table}[h]
\centering
\begin{tabular}{llllcccll}
\toprule
System & CV   &$dt$& $L$& $\tau_{\text{MetaD}}$ ($dt$)& $\tau_{\text{BDP}}$ ($dt$) & $\omega_{0}$& $\sigma_{G}$ & $\gamma$ \\
\midrule
MP& $x$ &0.001& 10& 100$L$& NA& 0.1& 0.08&20\\ 
Argon& $n_{\ell}$ &5 fs& 100& 100$L$& 20& 0.2& 0.5&8\\
HS&  $n_{c}$  &0.005&100& 500$L$& 20& 0.1& 2&10\\
DLVO&  $\xi$ &0.005&100& 500$L$& 20& 2& 1.3&40\\
BP& $Q_{6}$  &0.0005&100& 500$L$& 100&  1.0& 0.01&40\\
\bottomrule
\end{tabular}
\caption{hybridMC-MetaD simulation parameters for the systems studied in this work. CV is the collective variable. $L$ is the length of the short MD run in the hybridMC-MetaD simulation. $\tau_{\text{MetaD}}$ is the bias potential update interval. $\tau_{\text{BDP}}$ is the time constant for the BDP thermostat. $\omega_{0}$ and $\sigma_{G}$ are the initial height and width of the Gaussian bias potentials. $\gamma$ is the bias factor of well-tempered metadynamics.
}
\label{tab:params}
\end{table}

\subsubsection{Model potential system}
\label{sec:sim_setup_double_well}

We first constructed an illustrative model potential (MP) with two degrees of freedom $x$ and $y$. For simplicity, the $x$ and $y$ directions are designed to be completely separable. A potential barrier with a height of $2.0 \epsilon$ is constructed between two local minima at $(-1.0, 0.0)$ and $(1.0, 0.0)$ ($\epsilon$ is the reduced energy unit and default to $1$). The model potential can be formulated as follows:
\begin{align}
U(x,y)=B_0(x^2-x_0)^2 + K(y-y_0)^2,
\end{align}
where $B_0=2\epsilon$, $K=100\epsilon$, $x_0=1$ and $y_0=0$. For both unbiased and hybridMC-MetaD simulations, we initialized the system by placing a particle at $(1.0, 0.0)$ in a simulation box with an edge length of $5$. We then performed the simulation with the model potential using Brownian dynamics at $k_{B}T=0.2 \epsilon$, with an integration timestep of $dt = 0.001$. We ran $50$ independent unbiased simulations for $1 \times 10^{6}$ timesteps and $10$ independent hybridMC-MetaD simulations for $3 \times 10^{5}$ timesteps.

\subsubsection{Argon}
\label{sec:sim_setup_argon}

We study the condensation of argon at a moderate supersaturation level $S = 9.87$, for which we set the length of the cubic simulation box to $11 \text{nm}$. To model the interaction of the argon atoms, we use the Lennard-Jones pair potential \cite{chkonia2009evaluating} with $\epsilon = 0.99797$ kJ/mol and $\sigma = 0.3405$ nm, and truncate the potential at the cutoff \cite{tsai2019reaction} of 6.75$\sigma$. We apply a tail correction to the truncated potential such that it decays smoothly to zero at the cutoff.
At $S = 9.87$, we ran $60$ independent unbiased simulations and $25$ independent hybridMC-MetaD simulations. To generate a random initial configuration for each simulation, we first placed $N = 512$ argon atoms in a sparse cubic array, and thermalized the system at an elevated temperature $\text{T}_{0} = 1.2\text{T}$ for $2 \times 10^{5}$ timesteps (integration timestep $dt = 5$ fs). The temperature was then abruptly dropped to the target $\text{T} = 80.7 \text{K}$, and the system was further thermalized for $2 \times 10^{5}$ timesteps. Using the resulting system configurations, we ran the unbiased simulations for $3 \times 10^{7}$ timesteps, and the hybridMC-MetaD simulations for $4 \times 10^{7}$ timesteps.

\subsubsection{Nearly-hard bipyramids}
\label{sec:sim_setup_bipyramids}

The shape of a bipyramid (BP) can be described with two parameters \cite{lim2023engineering}: the number of edges in the cross-section polygon and the aspect ratio of the bipyramid, which is the ratio of the height of the bipyramid and the circumcircle diameter of the polygon. We ran simulations with N = 500 hexagonal bipyramids with an aspect ratio of $1.12$. The particles interact via the anisotropic Weeks-Chandler-Anderson pair potential (AWCA) potential \cite{ramasubramani2020mean}, and the resulting interaction is shape-dependent and effectively hard. We used the following parameters for the potential: $\epsilon = 0.1$, $\alpha = 0$ for repulsive contact-contact and center-center interaction, and $\sigma_{i} = \sigma_{i} = d$, where $d$ is the insphere diameter of the hexagonal bipyramid. The simulations were run at $k_{B}T = 1\epsilon$ and density $\rho\sigma^{3} = 0.545$. For each simulation, we generated a random initial configuration by placing the particles in a sparse cubic array at a much lower density and thermalizing the system for $5 \times 10^{5}$ timesteps (integration timestep $dt = 0.0005$). Then the simulation box was compressed to the target density ($\rho\sigma^{3} = 0.545$) over $5 \times 10^{5}$ timesteps. Random configurations at the target density were used for running $50$ independent unbiased simulations for $2 \times 10^{7}$ timesteps and $20$ independent hybridMC-MetaD simulations for $4 \times 10^{7}$ timesteps.

\subsubsection{Nearly-hard spheres}
\label{sec:sim_setup_hs}

We simulated the nearly-hard sphere (HS) system using the Weeks-Chandler-Andersen \cite{weeks1971role} (WCA) potential, which is the Lennard-Jones potential truncated at $r = 2^{1/6}\sigma$ and shifted smoothly to zero at the minimum. We ran simulations with N = 4,000 particles at $k_{B}T = 0.025\epsilon$ and densities $\rho\sigma^{3} = 0.778$ and $\rho\sigma^{3} = 0.762$. For both densities, we ran $50$ independent unbiased simulations and $20$ independent hybridMC-MetaD simulations. For all the simulations, we initialized the system by placing the particles in a sparse cubic array and setting the temperature to $k_{B}T = 0.3\epsilon$. The system was randomized for $2 \times 10^{5}$ timesteps (integration timestep $dt = 0.005$). Then the temperature was decreased to $k_{B}T = 0.025\epsilon$ and the system was thermalized for another $2 \times 10^{5}$ timesteps. From the resulting configurations, the unbiased simulations were run for $3 \times 10^{7}$ timesteps at $\rho\sigma^{3} = 0.762$, and $5 \times 10^{7}$ timesteps at  $\rho\sigma^{3} = 0.778$. The hybridMC-MetaD simulations were run for $5 \times 10^{7}$ timesteps at $\rho\sigma^{3} = 0.762$, and $4 \times 10^{7}$ timesteps at  $\rho\sigma^{3} = 0.778$.

\subsubsection{DLVO colloids}
\label{sec:sim_setup_dlvo}

To simulate a colloidal suspension, we use a DLVO pair potential with the parameters reported by Finney et al. \cite{finney2023variational}. We note that the potential's analytical form is different from the DLVO potential implemented in the simulation software we use, HOOMD-blue (version $4.7.0$), and we defined a table potential that has the same form as the one implemented in LAMMPS \cite{thompson2022lammps}.
Following Finney et al.'s approach, we studied a system of $N = 388$ particles at a number density of $\rho\sigma^{3} = 0.000485$. Instead of assigning the particles randomly to an FCC lattice, we placed them in a sparse cubic array in a cubic simulation box with periodic boundary conditions. The system was thermalized at an elevated temperature $k_{B}T_{0} = 1.2k_{B}T$ for $2 \times 10^{5}$ timesteps (integration timestep $dt = 0.005$) in each simulation. Then the temperature was decreased to the target $k_{B}T = 1.32 \epsilon$, and the system was further thermalized for $2 \times 10^{5}$ timesteps. Using these system configurations, we ran $50$ independent unbiased simulations for $3 \times 10^{8}$ timesteps, and $15$ independent hybridMC-MetaD simulations for $1 \times 10^{9}$ timesteps.

\subsection{Collective variables}
\label{sec:cv}

Next, we introduce the CVs we used for the systems studied in this work (also summarized in Table~\ref{tab:params}). For the model potential system, which only contains one particle, we directly biased the $x$-coordinate of the particle. For other systems, we employed more sophisticated CVs, and computed them with the open source particle simulation data analysis toolkit \href{https://freud.readthedocs.io/en/latest/}{\textit{freud}} \cite{ramasubramani2020freud} (version $3.0.0$).

\subsubsection{Number of liquid-like particles}

Following Tsai et al. \cite{tsai2019reaction}, we use the total number of liquid-like argon atoms ($n_{\ell}$) as the CV to describe argon liquid droplet nucleation from the gas phase. An atom is considered liquid-like if its coordination number is larger than a threshold value $c_{\ell}$, and the coordination number of atom $i$ is the number of its neighboring atoms within a radial cutoff $r_{c}$. In this case, we select an $r_{c} = 0.5$ such that $c_{i}$ is the coordination number of the first coordination shell. The total number of liquid-like atoms $n_{\ell}$ in the system can be then calculated using:
\begin{align}
    n_{\ell} = \sum_{i=1}^N \Theta(c_i - c_{\ell}),
    \label{eq:n}
\end{align}
where $\Theta$ is the Heaviside step function and $c_{\ell}=5$ is the threshold for considering an atom to be liquid-like \cite{ten1998computer}. We note that this order parameter is usually calculated using a switching function such that it is differentiable w.r.t. the positions of the atoms, but our version of $n_{\ell}$ is integer-valued and non-differentiable.

\subsubsection{Steinhardt bond-orientation order parameter}

To detect crystallization from the fluid phase in the nearly-hard bipyramid system, we use the Steinhardt order parameter \cite{steinhardt1983bond} $Q_\ell$ with $\ell=6$ as the CV. First, a quantity $q_{\ell m}(i)$ is calculated by summing the spherical harmonics of the angles $(\theta(\vec{r_{ij}}), \phi(\vec{r_{ij}}))$ between the vectors pointing from particle $i$ to its neighbors $j$:
\begin{align}
    q_{\ell m}(i)=\frac{1}{n_b(i)}\sum_{j=1}^{n_b(i)}Y_{\ell m}(\theta(\vec{r_{ij}}), \phi(\vec{r_{ij}})),
    \label{eq:q_lm}
\end{align}
where $n_b(i)$ denotes the total number of neighbors of particle $i$. Here we define particle $j$ as a neighbor of particle $i$ if their distance $r_{ij}$ is within a reduced distance of $r_{c}=2.0\sigma$. To yield a system-wide $Q_{\ell}$, the $q_{\ell m}$ values are averaged over all particles to get $\langle q_{\ell m} \rangle$, and Eq.~\ref{eq:q_l} is used to calculate the collective variable $Q_{\ell}$:
\begin{align}
    Q_\ell = \left[ \frac{4\pi}{2l+1}\sum_{m=-l}^{l}|\langle q_{\ell m} \rangle|^{2} \right]^{1/2}.
    \label{eq:q_l}
\end{align}

\subsubsection{Size of the largest solid cluster}

For the nearly-hard sphere system, we use the number of particles in the largest solid cluster ($n_{c}$) introduced by ten Wolde et al. \cite{wolde1996simulation} as our CV. This CV is also integer-valued and non-differentiable and thus is particularly useful for demonstrating the flexibility of our method. To calculate $n_c$, a bond orientational order parameter for each particle $i$ in the system is calculated between $i$ and its neighbor $j$ (Eq.~\ref{eq:q_lm}). Here the neighbor list of particle $i$ is constructed with a cutoff of $r_{c} = 1.5$. Then the following dot product is calculated:
\begin{align}
    d_\ell(i, j) = \frac{\sum_{m=-\ell}^\ell q_{\ell m}(i)q^{*}_{\ell m}(j)}{(\sum_{m=-\ell}^{\ell}|q_{\ell m}(i)|^{2})^{1/2}(\sum_{m=-\ell}^\ell |q_{\ell m}(j)|^{2})^{1/2}},
    \label{eq:d_ij}
\end{align}
where $l = 6$.
A bond is considered solid-like if $d_\ell(i, j) > d_{\rm threshold}$ (we use $d_{\rm threshold}=0.7$). If a particle has more than $\xi_c$ solid-like bonds ($\xi_c = 8$ in our case), the particle is considered solid-like.
A clustering algorithm is then performed on all the solid-like particles and the largest solid cluster in the system is identified. 

\subsubsection{Approximated reaction coordinate}

For the DLVO colloidal system, Meraz et al. \cite{meraz2024simulating} determined a reaction coordinate (RC) using the State-Predictive Information Bottleneck (SPIB) method \cite{wang2021state}. This RC is essentially a linear combination of the following order parameters: mean coordination number averaged over the entire system ($c$), the number of solid-like particles ($n_{s}$), the number of liquid-like particles ($n_{\ell}$), global Steinhardt order parameters $Q_{4}$ and $Q_{6}$, and the mean potential energy averaged over the system ($U$). Taking inspiration from their work,  we construct a CV ($\xi$) by linearly combining four of these order parameters using the same weights determined by Meraz et al.:
\begin{align}
    \xi = 0.3104 c + 0.1573 n_{\ell} + 0.4919 n_{s} + 0.2851U.
    \label{eq:spib}
\end{align}
Note that we do not include $Q_{4}$ or $Q_{6}$ in our CV as the weight of $Q_{4}$ indicates that its contribution to the RC is negligible \cite{meraz2024simulating}. We also found in our preliminary hybridMC-MetaD simulations that including $Q_{6}$ is not necessary. To calculate $n_{\ell}$ and $n_{s}$, we use a threshold of $c_{\ell} = 3$ for the liquid-like particles. The solid-like particles are identified using the same approach as the liquid-like particles but with a threshold of $c_{s} = 6$. The neighbor lists of the particles are constructed with a cutoff of $r_{c} = 6.4$.

\section{Results and Discussion}
\label{sec:results}


\subsection{Model potential}
\label{sec:test_model}

\begin{figure}[!t]
  \centering
  \includegraphics[width=8cm]{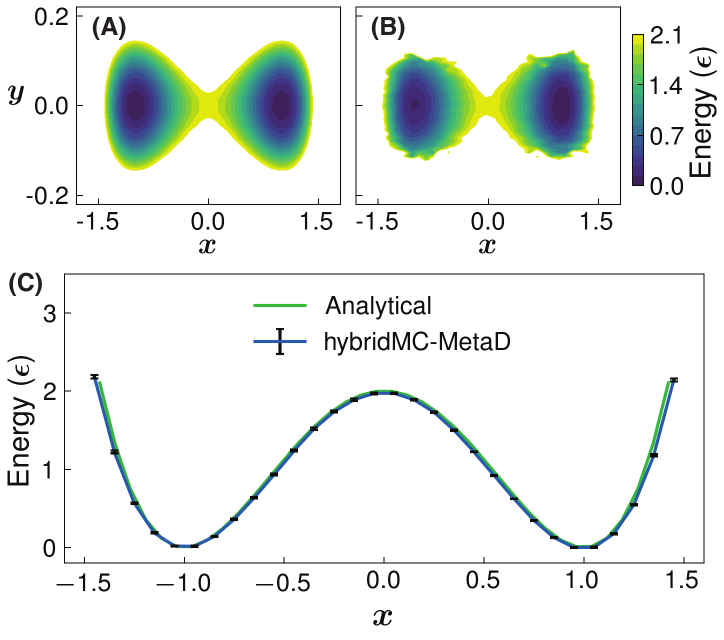}
  \caption
  {\textbf{Model potential} (A) Two-dimensional analytical potential energy surface and (B) energy surface reconstructed using data from a single hybridMC-MetaD simulation. (C) Free energy profile of the model potential system sampled using hybridMC-MetaD in comparison with the analytical potential. The hybridMC-MetaD free energy profile was constructed by averaging data of $10$ independent hybridMC-MetaD simulations. Standard errors are shown as error bars.}\label{fig:fe_model}
\end{figure}

At $k_{B}T=0.2 \epsilon$, the transition event across the barrier of the model potential is fairly rare, occurring only twice in $50$ independent unbiased simulations run for $1 \times 10^{6}$ timesteps. However, within $3 \times 10^{5}$ timesteps, the transition occurred in all $10$ independent hybridMC-MetaD simulations. In each hybridMC-MetaD simulation, we obtained a sufficient number of back-and-forth transitions between the local minima. Using one hybridMC-MetaD simulation trajectory, we calculated a free energy surface, shown in Fig.~\ref{fig:fe_model}(B). We also constructed a one dimensional free energy profile by averaging over the $10$ independent hybridMC-MetaD simulations (Fig.~\ref{fig:fe_model}(C)). Both the energy surface we sampled with hybridMC-MetaD and the reconstructed free energy profile match well with the analytical ground truth.
This simple case study provides an initial validation for our algorithm's accuracy.

\subsection{Argon liquid droplet nucleation}
\label{sec:test_argon}

\begin{figure}[!t]
  \centering

\includegraphics[width=8cm]{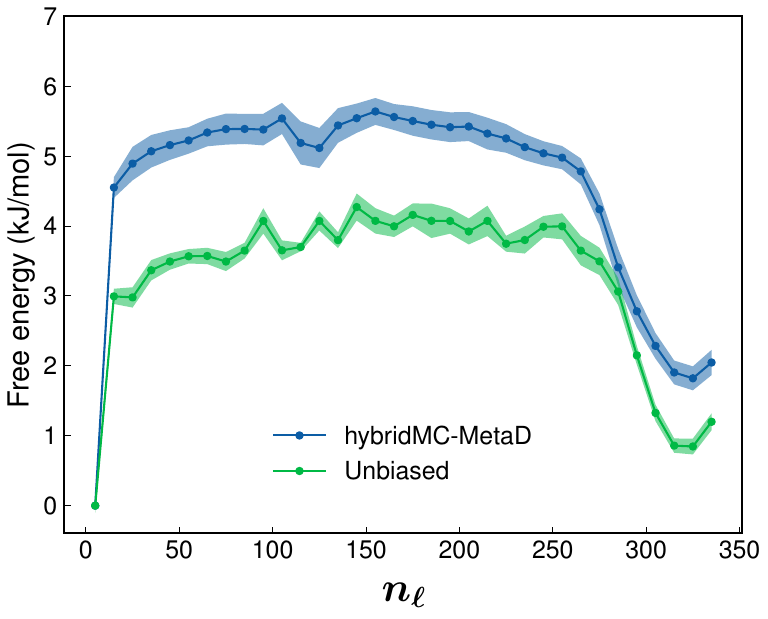}
  \caption
  {\textbf{Argon system} Free energy profiles along the number of liquid-like particles ($n_{\ell}$). The hybridMC-MetaD profile was constructed by reweighting and averaging over $25$ independent hybridMC-MetaD simulations, and the unbiased profile was calculated with data of $12$ independent unbiased simulations in which the system formed liquid droplets. The shading of the curves represents the standard error of the averaged results.
}\label{fig:fe_argon}
\end{figure}

In this section, we further validate our algorithm on a paradigmatic example: the nucleation of an argon liquid droplet \cite{salvalaglio2016overcoming, tsai2019reaction}. The underlying driving force for nucleation can be described by the supersaturation level $S$, the ratio of the actual vapor pressure and the equilibrium vapor pressure. We performed both unbiased simulations and hybridMC-MetaD simulations at a moderate supersaturation level $S = 9.87$, which has been reported \cite{tsai2019reaction} to have an energy barrier of $\Delta E (n)$ = $6.12$ kJ/mol, where $n$ is the number of liquid-like particles smoothed with switching functions such that it is continuous and differentiable.

At $S = 9.87$, we observed liquid droplet formation in only $12$ of the $60$ independent unbiased simulations, each run for $3 \times 10^{7}$ timesteps. However, within $1.0 \times 10^{7}$ timesteps,  nucleation of liquid droplets with $n_{\ell} \geq 100$ occurred in all $25$ independent hybridMC-MetaD simulations. Note that we did not have to transform the integer-valued number of liquid-like particles, $n_{\ell}$, into a differentiable one for the hybridMC-MetaD algorithm. We constructed two free energy profiles, one obtained by reweighting the hybridMC-MetaD simulation data, the other one with data from the $12$ unbiased simulations in which nucleation occurred (Fig.~\ref{fig:fe_argon}). The barrier of our reweighted free energy profile is around $5.83$ kJ/mol, in good agreement with Tsai et al.'s result \cite{tsai2019reaction} (differs by less than $0.5 k_{B}T$).
Even though the free energy profile constructed with unbiased simulation data shows a similar trend as the reweighted one, the barrier height is significantly lower. Such underestimation of the free energy barrier due to insufficient sampling has been previously discussed in Ref.~\citenum{meraz2024simulating}.

\subsection{Crystallization of nearly-hard spheres}
\label{sec:test_hs}

\begin{figure}[!t]
  \centering
  \includegraphics[width=8cm]{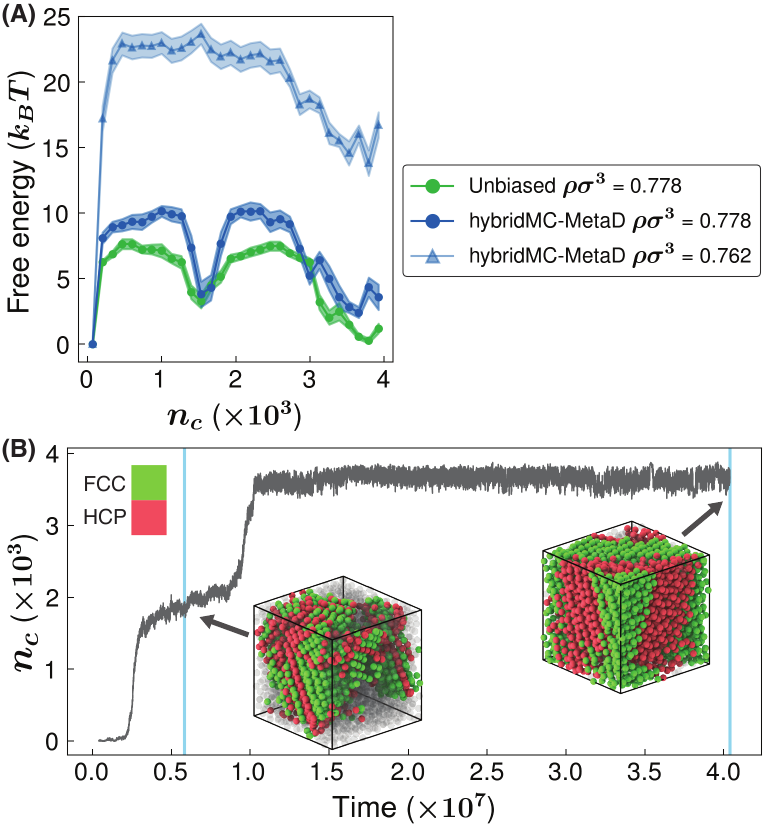}
  \caption
  {\textbf{Nearly-hard sphere system} (A) Free energy profiles along the number of particles in the largest solid cluster ($n_{c}$) obtained by averaging over the hybridMC-MetaD simulations at densities $\rho \sigma^{3} = 0.778$ ($20$ independent simulations) and $\rho \sigma^{3} = 0.762$ ($20$ independent simulations), and free energy profile along the same CV calculated by averaging over $43$ independent unbiased simulations at density $\rho \sigma^{3} = 0.778$. For all the profiles, we represent the standard error as error bars and the shading of the curves denotes the width of the error bars. (B) Two-step crystallization observed at $\rho \sigma^{3} = 0.778$ in both unbiased and biased simulations. Here we present a hybridMC-MetaD trajectory. Insets show solid clusters at around $n_{c} = 1700$ and $n_{c} = 3700$. The corresponding timesteps are labeled with sky blue vertical lines. We identified the local environments of particles using the Polyhedral Template Matching (PTM) method \cite{larsen2016robust} implemented in the visualization software OVITO \cite{stukowski2009visualization} and colored the particles accordingly. Noncrystalline particles are colored gray and rendered transparent for better visibility of crystalline particles.
}\label{fig:fe_hs}
\end{figure}

For a hard sphere system \cite{royall2024colloidal}, the particle interactions are purely entropic, and the equilibrium phase diagram depends on a single parameter, the packing fraction $\phi$. At low $\phi$, the system is thermodynamically stable in a fluid phase, whereas the crystalline phase is favored above $\phi = 0.54$, with freezing and melting packing fractions reported \cite{hunter2012physics} to be $\phi = 0.49$ and $\phi = 0.54$. 
In this section, we simulate a nearly-hard sphere system at two densities $\rho \sigma^{3} = 0.778$ and $\rho \sigma^{3} = 0.762$, which correspond to effective packing fractions \cite{filion2011simulation} $\phi_{\text{eff}} = 0.538$ and $\phi_{\text{eff}} = 0.526$. 

At $\rho \sigma ^3 = 0.778$ ($\phi_{\text{eff}} = 0.538$), we observed crystallization in $43$ of the $50$ independent unbiased simulations, each run for $5 \times 10^{7}$ timesteps. Out of the $43$ simulations, the system formed solid clusters with over $3000$ particles in $40$ simulations, and formed clusters with around $2000$ particles in $3$ simulations. We calculated a free energy profile using the $40$ unbiased trajectories with the larger clusters. In all $20$ independent hybridMC-MetaD simulations we ran at this density, the system formed solid clusters with over $3600$ particles. We also constructed a reweighted free energy profile using the hybridMC-MetaD data. Both free energy profiles show two barriers (Fig.~\ref{fig:fe_hs}(A)), and the barrier estimated from the unbiased trajectories is slightly lower than the reweighted one. As in the case study of the argon system, this is expected as we only accounted for the unbiased trajectories in which the system undergoes the entire crystallization process, and the sampling of the two barriers in the unbiased curve is insufficient.

To investigate the metastable state separating the barriers, we examined both the unbiased and biased simulation trajectories. We found that the state at around $n_{c} = 1500$ corresponds to a solid cluster coexisting with a fluid phase (Fig.~\ref{fig:fe_hs}(D)). This is consistent with the phase diagram of a perfectly hard sphere system, as the effective packing fraction of this density, $\phi_{\text{eff}} = 0.538$, is within the coexistence regime. It also explains why the system is trapped in solid clusters with $n_{c} \approx 2000$ in the $3$ unbiased simulations and our observation of two-step crystallization in both unbiased and biased simulations. An example is shown in Fig.~\ref{fig:fe_hs}(C). We note that the exact location of the metastable state may be affected by finite size effects; that is, the metastable state may appear at a larger $n_{c}$ if we ran larger simulations with more particles. Nonetheless, determining the exact $n_{c}$ of the metastable state is beyond the scope of this study, and the agreement in the trends of the two profiles further confirms the robustness of the hybridMC-MetaD algorithm.

At the lower density $\rho \sigma^{3} = 0.762$ ($\phi_{\text{eff}} = 0.526$), the system did not crystallize in any of the $50$ unbiased simulations within $3 \times 10^{7}$ timesteps. However, we accelerated crystallization significantly through hybridMC-MetaD. Solid clusters with at least $2000$ particles formed within $2 \times 10^{7}$ timesteps in all $20$ hybridMC-MetaD simulations, and the clusters reached sizes with $n_{c} \geq 3500$ in the end. The free energy profile reconstructed with hybridMC-MetaD simulation data is shown in Fig.~\ref{fig:fe_hs}(B). At this density, the barrier separating fluid and crystalline phases is approximately $24k_{B}T$, about $14k_{B}T$ higher than the barrier at $\rho \sigma^{3} = 0.778$, which explains the difficulty of observing crystallization in unbiased simulations.

\subsection{Two-step crystallization in a DLVO colloidal suspension}
\label{sec:test_dlvo}


\begin{figure*}[!t]
  \centering
  \includegraphics[width=17cm]{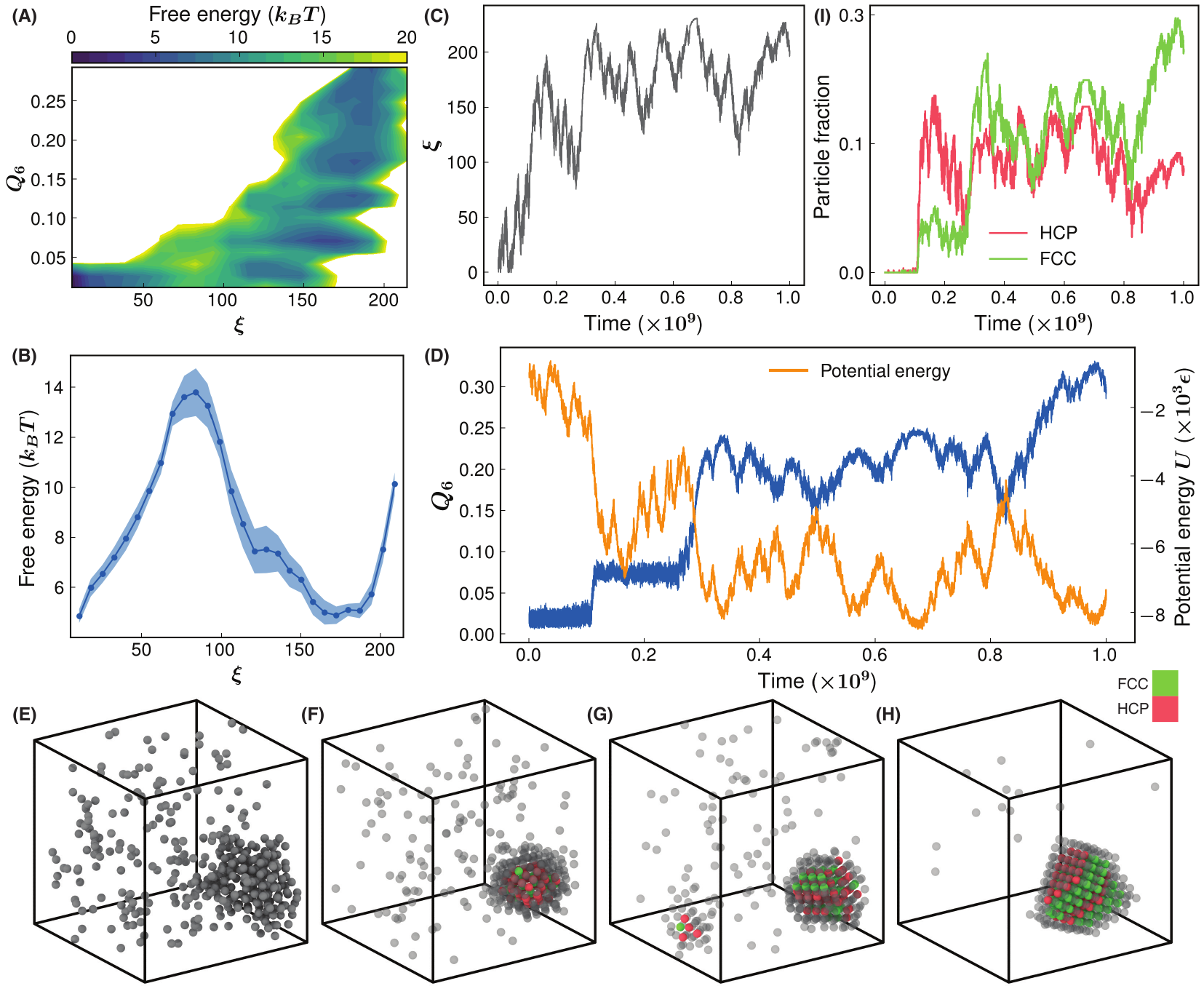}
  \caption
  {\textbf{DLVO system} (A) Two-dimensional free energy surface constructed as a function of $\xi$ (defined in Eq.~\ref{eq:spib}) and $Q_{6}$. (B) One-dimensional free energy profile along $\xi$ constructed by averaging over $15$ independent hybridMC-MetaD simulations. We represent the standard error as error bars, and the shading of the curve denotes the width of the error bars. Evolution of (C) $\xi$ and (D) $Q_{6}$ and potential energy of the entire system in one hybridMC-MetaD simulation. We show the following characteristic simulation configurations for the different stages during the two-step crystallization: (E) dense liquid droplet (timestep: $0.11 \times 10^{9}$, $Q_{6} = 0.049$, $U = -3105 \epsilon$), (F) cluster with low crystalline order (timestep: $0.20 \times 10^{9}$, $Q_{6} = 0.081$, $U = -5114\epsilon$), (G) cluster with intermediate crystalline order (timestep: $0.53 \times 10^{9}$, $Q_{6} = 0.185$, $U = -6464\epsilon$), and (H) crystal (timestep: $0.98 \times 10^{9}$, $Q_{6} = 0.328$, $U = -8232\epsilon$). Noncrystalline particles are colored gray in (E), and rendered transparent in (F-H). FCC-like particles are colored green and HCP-like particles are colored red in (F-H). Particles' local environments are identified with PTM. (I) Evolution of FCC-like and HCP-like particle fractions in the selected hybridMC-MetaD trajectory.
}\label{fig:fe_dlvo}
\end{figure*}

Next, we investigate the two-step crystallization of a colloidal suspension, a system first studied by Finney et al. \cite{finney2023variational}. Later, Meraz et al. \cite{meraz2024simulating} determined a reaction coordinate for the same system and explored the free energy landscape using well-tempered metadynamics. Both studies reported similar transitions: at $k_{B}T = 1.4 \epsilon$, the system initially formed a dense liquid droplet. In only $1\% - 2\%$ of their $1000$ unbiased simulations does the system continue to crystallize, and the system consistently crystallizes within the liquid droplet. Due to the different thermostats used (BDP in ours and velocity rescaling in Refs. \citenum{finney2023variational, meraz2024simulating}), we observed no transition at $k_{B}T > 1.3\epsilon$ within a reasonable number of timesteps in over $1000$ conventional MD simulations. However, for $1.32\epsilon \leq k_{B}T  \leq 1.4\epsilon$, by biasing $\xi$ (defined in Eq.~\ref{eq:spib}) in our hybridMC-MetaD simulations, we observed the same two-step crystallization. Here we focus on our results for $k_{B}T = 1.32 \epsilon$.

At $k_{B}T = 1.32\epsilon$, we ran $50$ independent unbiased simulations for $3 \times 10^{8}$ timesteps and observed no transitions at all. Throughout all the simulations, the potential energy of the system, $U$, stayed above $-2200 \epsilon$. Using the parameters reported in Table~\ref{tab:params}, we ran $15$ independent hybridMC-MetaD simulations, and in all $15$ simulations, the transitions were significantly accelerated: $U$ decreased to below $-3000 \epsilon$ within $2.5 \times 10^{8}$ timesteps, at which point the system had already formed a liquid droplet at least once. To quantify the structural order in the system, we computed $Q_{6}$ for the hybridMC-MetaD trajectories using Eq.~\ref{eq:q_l}. In this case, we used a fixed number of neighbors, $n_{b}(i) = 12$, instead of a cutoff radius to build the neighbor lists. This is necessary because after liquid droplet formation, the particles still in the vapor phase have zero neighbors within a smaller fixed distance. With no neighbors, calculation of the Steinhardt order parameter breaks down. The evolution of $Q_{6}$ and $U$ of a single biased simulation trajectory is shown in Fig.~\ref{fig:fe_dlvo}(D). The corresponding $\xi$ is shown in Fig.~\ref{fig:fe_dlvo}(C). We constructed a two-dimensional free energy surface as a function of $\xi$ and $Q_{6}$ by reweighting the hybridMC-MetaD simulation data (Fig.~\ref{fig:fe_dlvo}(A)), and calculated a one-dimensional free energy profile along $\xi$ (Fig.~\ref{fig:fe_dlvo}(B)). As can be seen from the free energy surface and the evolution of $Q_{6}$, by biasing $\xi$, we also enhanced the fluctuations in $Q_{6}$. On the two-dimensional free energy surface, the two steps of the crystallization process appear to be almost orthogonal, with $\xi$ increasing first. In addition to the initial vapor phase, the system has $4$ other metastable states, separated by multiple barriers. The energy barrier along $\xi$ (around $14 k_{B}T$, Fig.~\ref{fig:fe_dlvo}(B)) is the highest one.

For each of the metastable states, a characteristic simulation configuration is shown in Fig.~\ref{fig:fe_dlvo}(E-H). The system first forms a dense liquid droplet with $Q_{6} < 0.05$ (Fig.~\ref{fig:fe_dlvo}(E)). Even though the structural order of this liquid phase is still low, and it cannot be distinguished from the vapor phase with $Q_{6}$, particles in the liquid droplet have higher coordination numbers and hence higher $\xi$. Next, within the liquid droplet, crystalline order starts to develop. At the early stage of crystallization, most of the crystalline particles are HCP-like (Fig.~\ref{fig:fe_dlvo}(F)), and many particles are still in the vapor phase. Then more and more particles attach to the crystal cluster and the number of FCC-like particles within the cluster increases, accompanied by an increase in $Q_{6}$ (Fig.~\ref{fig:fe_dlvo}(G)). Finally, most of the particles in the system become part of the crystal cluster, and FCC-like particles outnumber HCP-like ones (Fig.~\ref{fig:fe_dlvo}(I)), with $Q_{6}$ reaching $0.3$. We note that the trends in particle fractions are specific to the trajectory shown here.
However, the formation of clusters with increasing crystalline order within the liquid droplet was observed in all our hybridMC-MetaD simulations, and the competition between FCC- and HCP-like particles explains the many local minima in the rugged free energy landscape.

\subsection{Self-assembly of nearly-hard bipyramids}
\label{sec:test_hard_polyhedra}

Finally, we consider crystallization in a dense system of hard shapes.
Lim et al. \cite{lim2023engineering} reported that a family of hard bipyramids can self-assemble into a plethora of mesophases, i.e., liquid crystals or plastic crystals, some of which continue to undergo transitions to eventually form both orientationally and translationally ordered crystals. Here we investigate the self-assembly of a system of one such bipyramid shape, a hexagonal bipyramid with an aspect ratio of $1.12$. As shown in Ref. \citenum{lim2023engineering}, this bipyramid system forms a plastic BCC (pBCC) crystal at intermediate pressures. Upon further compression, the system develops orientational order and transitions to a body-centered tetragonal (BCT) crystal at reduced pressure $P^{*} = 17.0$.

\begin{figure}[!t]
  \centering
  \includegraphics[width=8.04cm]{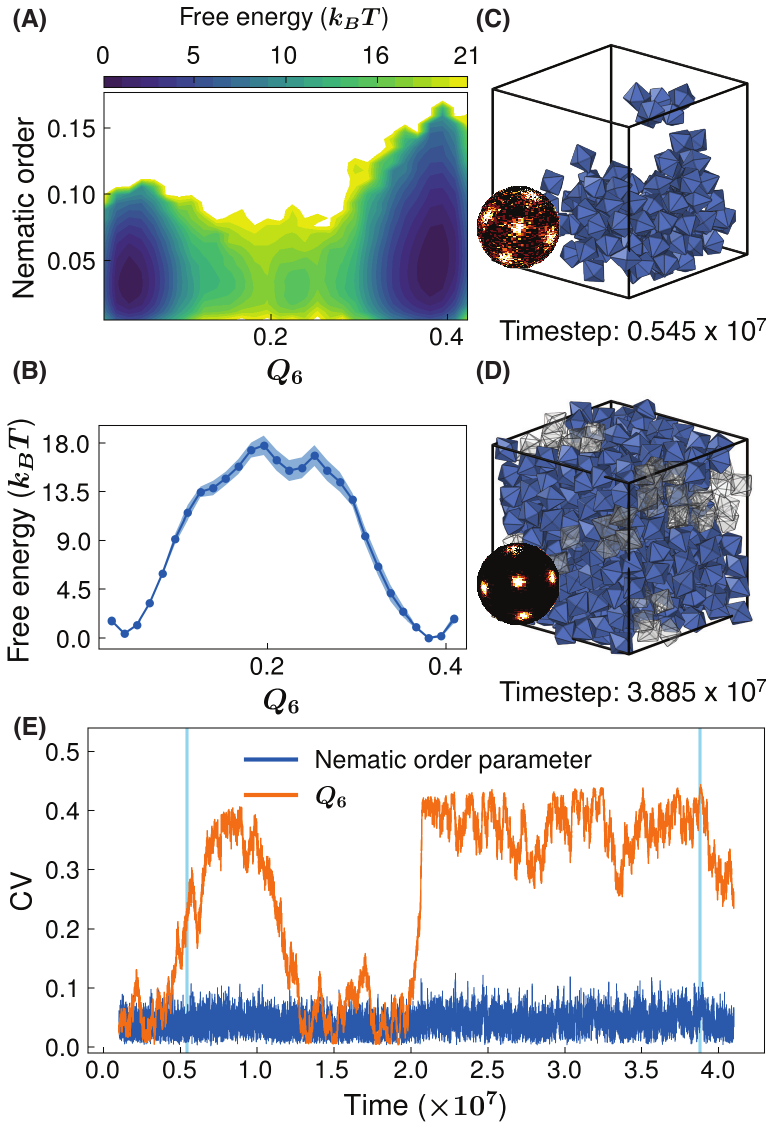}
  
  \caption
  {\textbf{Nearly-hard bipyramid system} (A) Free energy surface constructed as a function of $Q_{6}$ and the nematic order parameter, using data from $20$ independent hybridMC-MetaD simulation trajectories. (B) One-dimensional free energy profile along $Q_{6}$ constructed by averaging over $20$ independent hybridMC-MetaD simulations. We represent the standard error as error bars and the shading of the curve denotes the width of the error bars. We also present simulation snapshots of the system corresponding to (C) $Q_{6} = 0.2$ and (D) $Q_{6} = 0.4$. The bond orientational order diagrams are shown as insets, whose patterns are indicative of the BCC crystalline order. We identified the local environments of particles using PTM, and the BCC-like particles are colored blue. Noncrystalline particles are not shown in (C) and are rendered transparent for better visibility of the crystalline particles in (D). (E) $Q_{6}$ (directly biased CV) and nematic order parameter of one hybridMC-MetaD simulation. $Q_{6}$ shows the characteristic enhanced fluctuations and a back-and-forth transition between the disordered state and the pBCC crystal. As the system does not form the BCT crystal at our density of interest, the nematic order remains low.
}\label{fig:fe_bp}
\end{figure}

At $\rho\sigma^{3} = 0.545$, we observed no crystallization in the $50$ independent unbiased simulations run for $2 \times 10^{7}$ timesteps. However, using the parameters listed in Table~\ref{tab:params}, we significantly accelerated the fluid $\rightarrow$ pBCC transition with hybridMC-MetaD by biasing $Q_6$ alone. In all $20$ independent hybridMC-MetaD simulations, the system reached $Q_{6} = 0.2$ within $1.5 \times 10^{7}$ timesteps, which corresponds to a large pBCC cluster, as shown in Fig.~\ref{fig:fe_bp}(C). Furthermore, in $5$ of our hybridMC-MetaD simulations, back-and-forth transitions between the fluid phase and the pBCC crystal were observed. An example is shown in Fig.~\ref{fig:fe_bp}(E).

We constructed a two-dimensional free energy surface as a function of $Q_{6}$ and the nematic order parameter and calculated a one-dimensional free energy profile along $Q_{6}$ (Fig.~\ref{fig:fe_bp}(A, B)).
As the pBCC $\rightarrow$ BCT transition occurs at reduced pressure $P^{*} = 17.0$, which corresponds to a packing fraction of roughly $0.6$, we do not expect to observe this transition at our density of interest ($\rho\sigma^{3} = 0.545$). 
Therefore, as shown in Fig.~\ref{fig:fe_bp}(E), 
the nematic order parameter remains low throughout the biased simulations, despite the increase in $Q_{6}$. 
This observation is consistent with the L-shaped pathway \cite{lim2023engineering} reported by Lim et al., along which only $Q_6$ increased until the reduced pressure reached $P^{*} = 17.0$. 

As shown in Fig.~\ref{fig:fe_bp}(A, B), the two local minima, fluid and pBCC, are separated by one barrier along $Q_{6}$ with a height of around $18 k_{B}T$. The high barrier explains why no crystallization is observed in any of the unbiased simulations. In addition, as can be seen from Fig.~\ref{fig:fe_bp}(A), the local minimum corresponding to the pBCC phase is more spread out towards higher nematic order values, indicating that the pBCC crystal can reach slightly higher nematic order than the fluid phase at this density ($\rho\sigma^{3} = 0.545$).

\section*{Conclusions}
\label{sec:conclusion}

In summary, we proposed a new enhanced sampling scheme which we term hybridMC-MetaD (Hybrid Monte Carlo Metadynamics). 
By virtue of the Metropolis criterion in the hybridMC-MetaD algorithm, we can eliminate the evaluation of bias forces (and bias torques for non-spherical particles) and the calculation of derivatives of the CVs. 
In this way, not only does our algorithm enable the usage of a wider range of CVs, such as those that are non-differentiable, it also simplifies the execution of metadynamics simulations for anisotropic bodies, such as colloidal nanoparticles. 

We demonstrated the versatility of our method by probing rare events in several different systems, including the rare transition across the energy barrier in an illustrative model potential system, liquid-droplet formation in the argon system, crystallization processes in a nearly-hard sphere system, a DLVO colloidal suspension and a nearly-hard bipyramid system. For all these systems, we utilized non-differentiable CVs in our hybridMC-MetaD simulations (either the CV itself is integer-valued, or the neighbor lists used to construct the CVs are discontinuous), and observed significant acceleration and much-improved sampling of the transitions. Therefore, our hybridMC-MetaD algorithm provides an alternate approach to using metadynamics with conventional MD simulations, offering users greater flexibility in choosing and designing suitable CVs. With an easier and more user-friendly implementation of the biasing scheme, we believe our method paves the way for broader applications of the powerful metadynamics method.

\section*{Conflict of Interest}
\label{sec:conflicts_of_interest}
The authors declare no conflict of interest.


\section*{Acknowledgments}
\label{sec:acknowledgements}

This research was supported by a Computational and Data-Enabled Science and Engineering (CDS\&E) grant from the U.S. National Science Foundation, Division of Materials Research Award No. DMR 2302470.  This work used SDSC Expanse and Purdue Anvil through allocation DMR 140129 from the Advanced Cyberinfrastructure Coordination Ecosystem: Services \& Support (ACCESS) program, which is supported by National Science Foundation grants No. 2138259, No. 2138286, No. 2138307, No. 2137603, and No. 2138296. Computational resources and services were also provided by Advanced Research Computing at the University of Michigan, Ann Arbor. The authors also thank Dr. Joshua A. Anderson and Shih-Kuang Alex Lee for helpful discussions.

\section*{Data Availability}
\label{sec:dataavailability}
Code to implement the hybridMC-MetaD algorithm is available at \url{https://github.com/Charlottez112/hybridMC-MetaD}, along with scripts for running both unbiased and hybridMC-MetaD simulations for the double-well potential system. We also include our implementation of the algorithm as a HOOMD-blue custom updater in the same repository, with which we performed all of our simulations. The source code and the data generated for this study are in the process of being deposited in the \href{https://deepblue.lib.umich.edu/data}{Deep Blue Data repository} and will be publicly accessible upon completion. 

\appendix*
\section{Derivation of the Acceptance Criterion for the hybridMC-MetaD Algorithm}
\label{app:derivation}
To ensure detailed balance for our hybridMC-MetaD algorithm, the following condition must be fulfilled: $\pi_{\rm bias}(\mathbf{x})q(\mathbf{x}\to\mathbf{x}')=\pi_{\rm bias}(\mathbf{x}')q(\mathbf{x}'\to\mathbf{x})$, where $\mathbf{x}$ denotes a system configuration, $\pi_{\rm bias}(\mathbf{x})$ is the biased equilibrium distribution of $\mathbf{x}$, and $q(\mathbf{x}\to\mathbf{x}')$ is the probability for the system to transition from $\mathbf{x}$ to a new configuration $\mathbf{x}'$.
If we propose a new system configuration by performing a short NVT or canonical MD simulation, we can rewrite the left-hand side of the detailed balance as follows:

\begin{equation}
\begin{aligned}
    \pi_{\rm bias}(\mathbf{x})q(\mathbf{x}\to\mathbf{x}') = e^{-\beta V(\mathbf{x})}\alpha(\mathbf{x}\to\mathbf{x}') \\
    \int\int d\mathbf{p}_0d\mathbf{p}_t \pi(\mathbf{x}_0,\mathbf{p}_0)P(\mathbf{x}_0,\mathbf{p}_0 \to \mathbf{x}_t,\mathbf{p}_t), \label{eq:forward}
\end{aligned}
\end{equation}
where $V$ is the bias potential, $\alpha(\mathbf{x}\to\mathbf{x}')$ is the acceptance probability, $\mathbf{p}$ denotes particle momenta, $\pi$ is the unbiased equilibrium distribution of system configurations and $P$ is the proposal probability. $\mathbf{x}_0 = \mathbf{x}$ is the initial system configuration, and $\mathbf{x}_t = \mathbf{x}'$ is the new configuration obtained at time $t$.
The right-hand side of the detailed balance can be written as:
\begin{equation}
\begin{aligned}
    \pi_{\rm bias}(\mathbf{x}')q(\mathbf{x}'\to\mathbf{x}) = e^{-\beta V(\mathbf{x}')}\alpha(\mathbf{x}'\to\mathbf{x}) \\
    \int\int d\mathbf{p}_0d\mathbf{p}_t \pi(\mathbf{x}_t,\mathbf{p}_t)P(\mathbf{x}_t,\mathbf{p}_t \to \mathbf{x}_0,\mathbf{p}_0). \label{eq:old_backward} \\
\end{aligned}
\end{equation}

As described in Ref.\citenum{van1992stochastic}, a system that is supplied with a heat bath and is in the canonical ensemble can be treated as a stationary Markov chain and satisfies the following identity:
\begin{equation}
\begin{aligned}
    & \pi(\mathbf{x},\mathbf{p}) P(\mathbf{x}, \mathbf{p} \to \mathbf{x}',\mathbf{p}')\\
    & = \pi(\mathbf{x}',-\mathbf{p}') P(\mathbf{x}',-\mathbf{p}' \to \mathbf{x},-\mathbf{p}).
    \label{eq:id_stationary_MC}
\end{aligned}
\end{equation}

Using this identity, we can rewrite Eq.~\ref{eq:old_backward}:
\begin{equation}
\begin{aligned}
    &e^{-\beta V(\mathbf{x}')}\alpha(\mathbf{x}'\to\mathbf{x}) \\
    &\int\int d\mathbf{p}_0d\mathbf{p}_t \pi(\mathbf{x}_t,\mathbf{p}_t)P(\mathbf{x}_t,\mathbf{p}_t \to \mathbf{x}_0,\mathbf{p}_0) \\
    &=e^{-\beta V(\mathbf{x}')}\alpha(\mathbf{x}'\to\mathbf{x}) \\
    &\int\int d(-\mathbf{p}_0)d(-\mathbf{p}_t) 
    \pi(\mathbf{x}_t,-\mathbf{p}_t)P(\mathbf{x}_t,-\mathbf{p}_t \to \mathbf{x}_0,-\mathbf{p}_0) \\
    &=e^{-\beta V(\mathbf{x}')}\alpha(\mathbf{x}'\to\mathbf{x}) \\
    &\int\int d\mathbf{p}_0d\mathbf{p}_t \pi(\mathbf{x}_0,\mathbf{p}_0)P(\mathbf{x}_0,\mathbf{p}_0 \to \mathbf{x}_t,\mathbf{p}_t). \label{eq:backward}
\end{aligned}
\end{equation}
Comparing Eq.~\ref{eq:forward} and Eq.~\ref{eq:backward}, we obtain:
\begin{align}
    \frac{ \alpha(\mathbf{x}\to\mathbf{x}') }{ \alpha(\mathbf{x}'\to\mathbf{x}) }
    =\frac{e^{-\beta V(\mathbf{x}')}}{e^{-\beta V(\mathbf{x})}}.
\end{align}
Therefore, the modified Metropolis acceptance criterion that satisfies the detailed balance condition is
\begin{align}
    \alpha(\mathbf{x}\to\mathbf{x}')=\min\left[1,\frac{e^{-\beta V(\mathbf{x}')}}{e^{-\beta V(\mathbf{x})}}\right],
    \quad
    \alpha(\mathbf{x}'\to\mathbf{x})=1.
\end{align}

\bibliography{references.bib}

\end{document}